\documentclass[aps,nofootinbib,longbibliography,prd,twocolumn]{revtex4-1}
\usepackage[babel]{csquotes}
\usepackage{graphicx}
\usepackage{amsmath,amssymb}
\usepackage[colorlinks,citecolor=blue,linkcolor=blue,urlcolor=blue]{hyperref}
\usepackage{mathrsfs}
\usepackage{enumerate}
\def\be{\begin{equation}}
\def\ee{\end{equation}}

\begin{document}

\title{Where was past low-entropy?}

\author{Carlo Rovelli${}^b$ \vspace{1mm}}

\affiliation{\small\mbox{CPT, Aix-Marseille Universit\'e, Universit\'e de Toulon, CNRS, F-13288 Marseille, France.}
}
\date{\small\today}

\begin{abstract}

\noindent 
I identify where precisely is low-entropy located in the early universe, arguing that some common answers overplay the role of gravity. I discuss the available  interpretations of the ``improbability" of this early low-entropy.  
\end{abstract}

\maketitle

\section{Introduction}

The first part of this note traces the source of physical irreversibility, identifying \emph{where} precisely is low-entropy  located in the early universe. I argue against some common answers that overplay the role of gravity (see also \cite{Earman2006}).  The issue stems from the fact that the  standard model of cosmology \cite{Weinberg2008} assumes matter to be at equilibrium in the early universe --hence at maximal entropy--, in apparent tension with the past low-entropy needed to understand the ubiquitous irreversibility of macroscopic phenomena \cite{Lebowitz1993,Albert2000,Price}.  Key observations  in this first part of the paper come from \cite{Wallace2010}.  The conclusions of this first part seem solid to me. 

The second part of this note discusses the remaining open question regarding irreversibility: can we make sense of the  ``improbability" or ``atypicality"  of this early low entropy? I discuss tentative solutions, including the possibility of a perspectival interpretation of entropy \cite{Rovelli2017c}.

\section{Part I.  Where?}

\subsection{A simple model, first version}

Let me start my reminding the reader of some well-established facts in thermodynamics. These are illustrated by a simple model that captures some salient aspects of the thermal history of the universe.  

\begin{figure}[b]
\includegraphics[height=2.5cm]{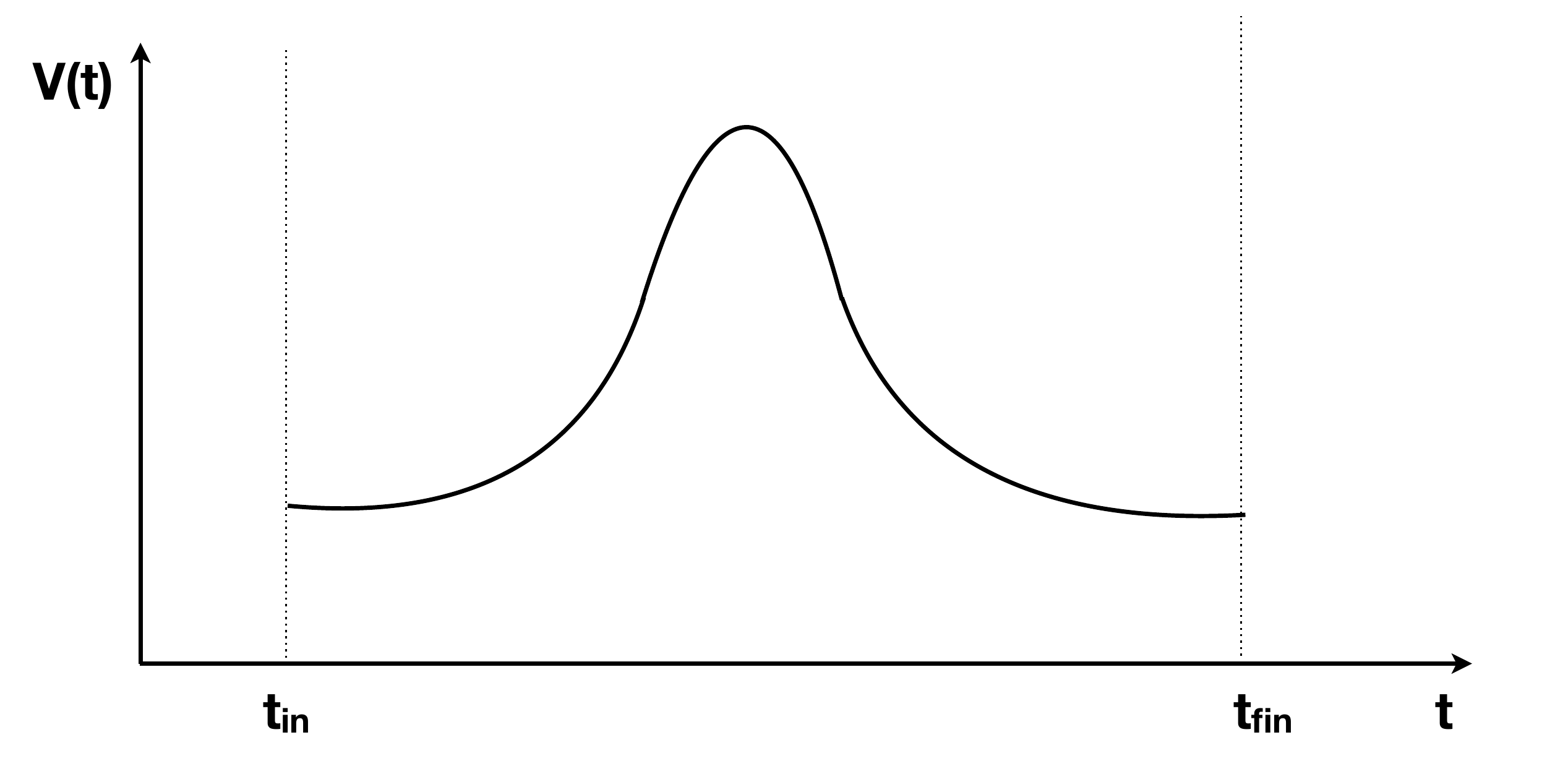}
\caption{Time evolution of the volume in the gas model.}
\label{uno}
\end{figure}

Consider an ideal gas in a thermally isolated container whose volume $V$ can be modified with a piston, moved by an external force.    Assume that the volume  increases from the initial value $V$ at some negative time $t_{0}=-t_{f\!in}$, to a maximum value $V_{max}$ at $t=0$ and then symmetrically decreases back to $V$ at positive time $t_{f\!in}$, so that the volume evolves in time respecting $V(t)=V(-t)$  as in Figure \ref{uno}.  

Assume that at  time $t_{0}$ the gas was in thermal equilibrium at temperature $T$.   Does the entropy of the gas increase in the process?  

The answer depends on the rapidity of the volume change.  If this is much slower than the thermalization time of the gas, the gas remains constantly close to equilibrium. From $V(t)=V(-t)$ and the equilibrium relation $pV=constant$ it follows that the pressure $p$ satisfies  $p(t)=p(-t)$, and therefore the work $W=\int p\, dV$ done on the gas during the compression is equal to the work extracted from the gas during the expansion. Hence the net exchange of energy is zero. At $t_{f\!in}$ the gas will therefore have the same temperature as the initial one, hence the same entropy, and there is no increase of entropy in the process. The process is reversible.  
 
\begin{figure}[t]
\raisebox{.3cm}{\includegraphics[height=1.4cm]{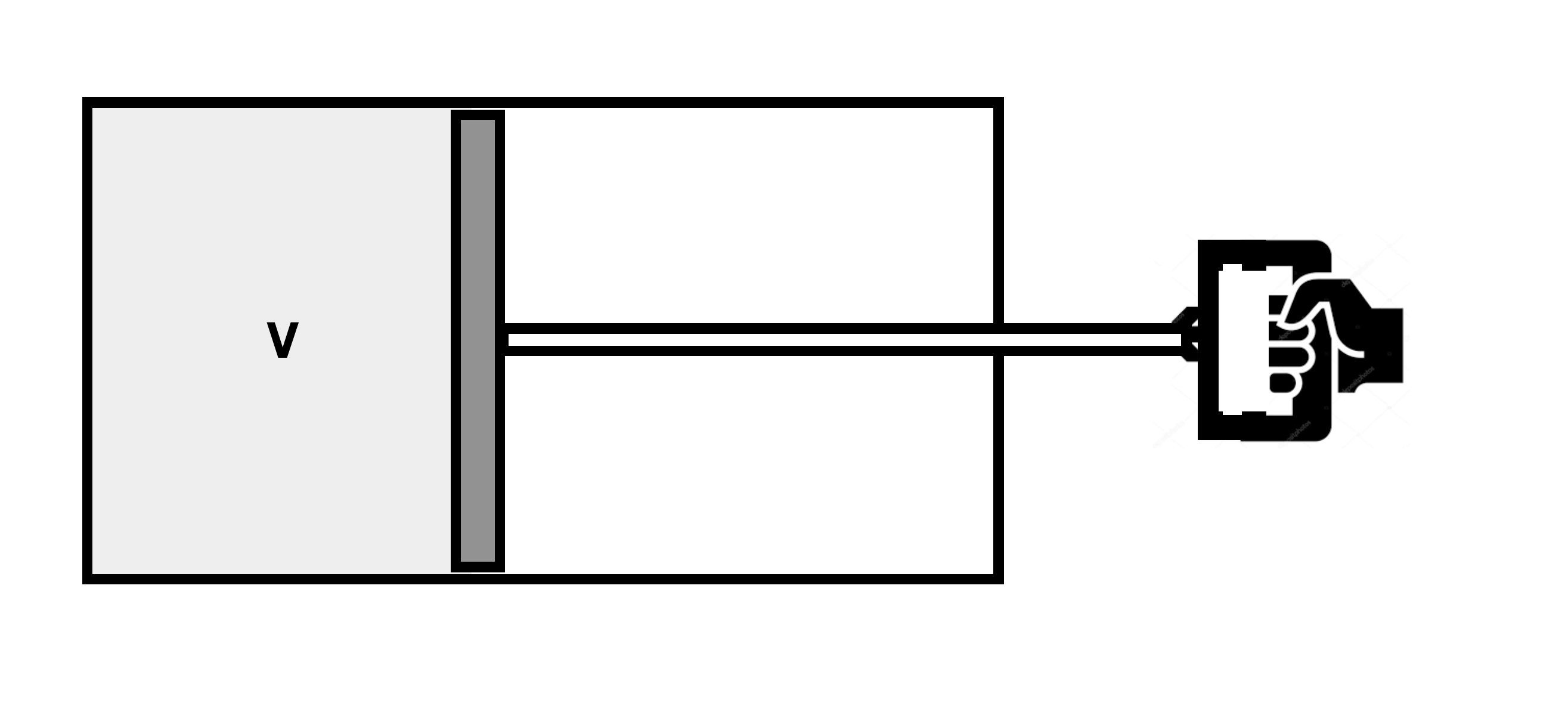}}\hspace{1cm}\includegraphics[height=2cm]{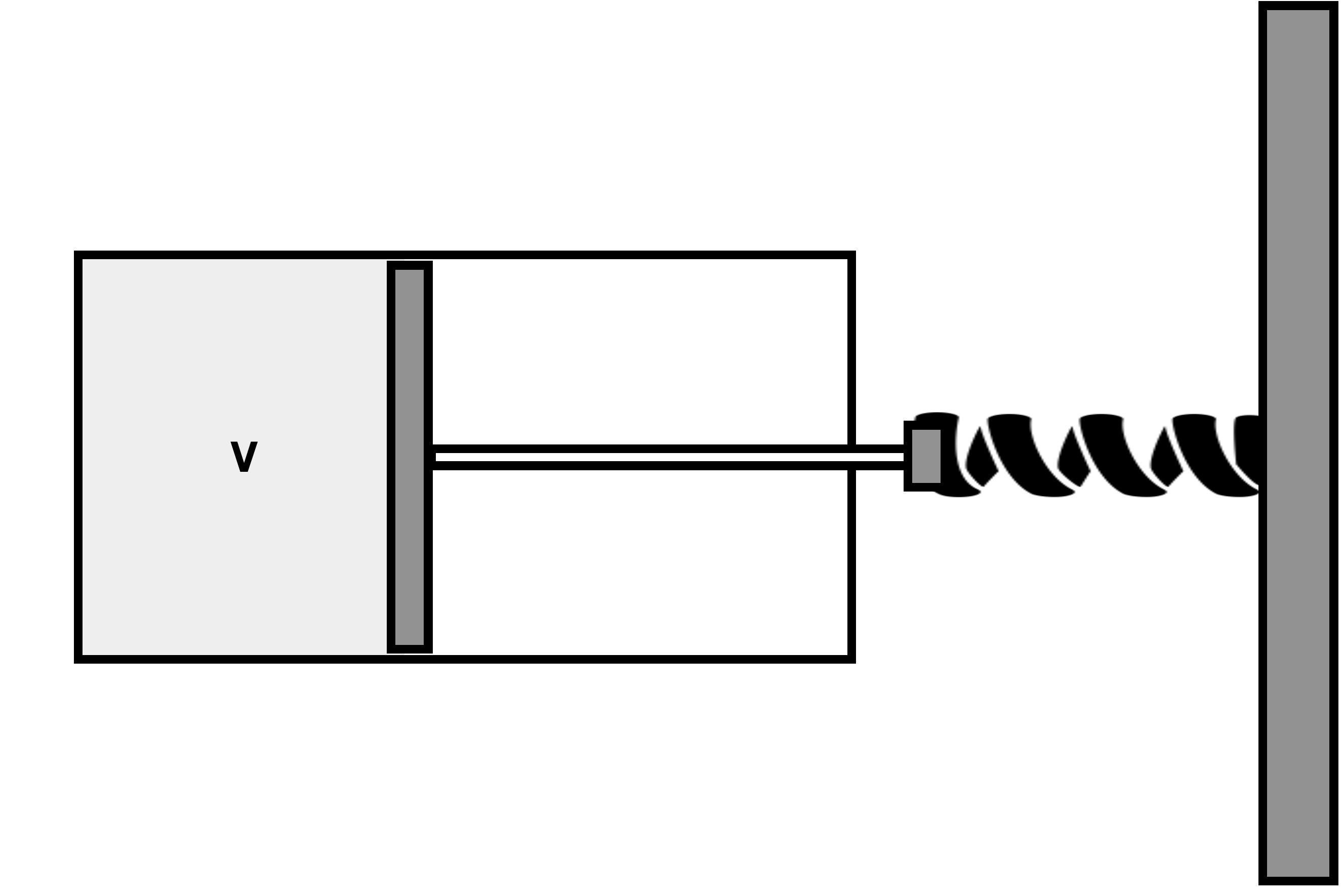}
\caption{The two models considered in the paper. Left: A system formed by the gas in a chamber; its volume is changed by a piston moved by an external force. Right: A system formed by a gas in a chamber plus the piston attached to a spring.}
\label{due}
\end{figure}

But if the volume changes fast enough, entropy increases and the process is irreversible: temperature at  $t_{f\!in}$ is higher than at $t_{0}$.  Mechanically, this is easy understood: during the fast expansion, there are less molecules that hit the piston than during the fast compression. In the limit case where the speed of the piston is much faster than molecular speeds, the pressure during expansion vanishes, hence no work is extracted during the expansion, while work is done in the compression.   Since the work during compression is larger than the work extracted during expansion, energy is put into the gas, and the final temperature is higher than the initial one.  The process is irreversible.  During the process, the gas finds itself away from equilibrium. 

The moral is that a gas initially at equilibrium in a space that expands faster than its thermalisation time goes out of equilibrium and generates irreversibility.  

As illustrated below, this is precisely what has happened to the universe in which we live: its rapid cosmological expansion has driven its matter content out of equilibrium.   \emph{This} is by far the main source of the present irreversibility in the cosmos, as I discuss more in detail below.  Before doing so, however, we need to refine the model. 

\subsection{A simple model, second version}

The model discussed above includes an external force that drives the piston.  There are no external forces in the universe: all degrees of freedom interact with one another and belong to the same coupled dynamics. 

Among these degrees of freedom is the cosmological phase factor $a(t)$, which is the dynamical variable that governs the available volume to each 
co-mouving region of space in a Friedman cosmology 
\be
ds^2=-dt^2+a^2(t)\,d\vec x^2,
\ee
where $d\vec x^2$ is the metric of a homogenous 3d space. The single degree of freedom $a(t)$ acts very much like the volume $V(t)$ of the above model, but it is not driven by external forces: it is driven by a dynamical interaction with all the other the degrees of freedom of the universe.   To better mimic the thermal history of the universe, let us therefore modify the model above and get rid of the external force.   

Consider again a gas in a volume with a piston, but assume that now the piston attached to a spring, and free to interact and exchange energy with the gas.  See Figure \ref{due}. 

Let us denote $X(t)$ the position of the piston at time $t$.  With appropriate parameters for the dynamics of the piston-spring system and appropriate initial conditions $X(t_{0}),\dot X(t_{0})$, the variation of the volume in time between $t_{0}$ and $t_{fin}$ can be set to be similar to the previous case. 

If we disregard dissipation, the piston continues to oscillate indefinitely.   But since the temperature and the entropy of the gas increase at each oscillation, there must be dissipation. If we take the dissipation in the gas into account, the force $F= \int p\; dA$ that the gas exerts on the piston during expansion is lower than during  compression. Therefore the piston/spring system gradually transfers the (kinetic and potential) energy of the piston to the gas. The temperature and the entropy of the gas raise.  The process is irreversible. 

The source of the irreversibility is transparent: since there is no external force , we  interpret irreversibility as simple redistribution of energy.  

In fact, if we let the system continue to evolve freely, the piston will repeat the oscillation many times.  At each oscillation, the temperature of the gas raises, at the expense of the energy in the piston-spring system.  The oscillations slowly reduce in amplitude until the piston sets down to its equilibrium position $X_o$.  

More precisely, the piston does not set down to its exact mechanical equilibrium position, but it rather fluctuates thermally around this position, since it is coupled to the hot gas.  At this point thermal equilibrium is achieved, the temperature of the gas stops increasing, and the entropy has reached its maximum. 
The source of irreversibility in this entire process is clear: although the gas was in equilibrium \emph{by itself}, the system gas+piston/spring was far from equilibrium.   There was far more energy in the piston/spring system than its proper equilibrium share.  

The system as a whole was badly out of equilibrium, hence irreversibility.  Where was the low entropy? What was out of equilibrium? Clearly the answer is the piston/spring system, namely the single variable $X(t)$.  A single variable far away from its equilibrium value is sufficient to have very low entropy because it can absorb a big share of energy leaving the rest of the variables with low energy and hence with a much smaller available phase space. That is: low entropy. 
 
This is what has happened in our universe. (In Appendix \ref{A1} I give a simple toy model better illustrating this fact in the context of the Friedman equation,)
 The state of the system shortly after the big bang was badly out of equilibrium because one single degree of freedom, the scale factor $a(t)$, was.   The entire initial low-entropy of the universe is entirely stored in the out-of-equilibrium value of a single variable.   In the following section, I make this claim concrete. 

\subsection{The thermal history of the universe}

To understand the thermal history of the universe, the notion of \emph{metastable}, or quasi-equilibrium state, and the related notion of  \emph{channel} for entropy-increase, are essential.   

A pile of wood in a room full of air is quite thermodynamically stable: it can remain as it is for many years.   But the basic elements forming it are not at the maximum value of their entropy. Far from that.  This is obvious from the fact that if we ignite a fire, the wood burns.  Burning is a violent irreversible phenomenon that dramatically increases entropy.  After burning, the content of the room is reduced to ashes smoke and vapour, which form a much higher-entropy state of the ingredients in the room, than the initial wood and air.   Therefore the constituents forming the pile of wood in the room full of air are in a remarkably stable configuration and yet far from their maximum entropy state.  They are in a metastable state, or quasi-equilibrium state.  

The reason metastable state exist is that there can be obstructions in the phase space of a system, which do not allow the system to easily explore its entire phase space.   The system remains trapped in a relatively small region of phase space for a very long time.  In other words, the scale of its thermalisation time can be very long.   This may change if some dynamical event allows the system to overcome the obstruction, thus opening a \emph{channel} through which the system can exit the phase space region where it was trapped, and move out to a larger region of its phase space.  In the example above, the channel is represented by the combustion process and is opened by igniting the fire. 

Metastable states are ubiquitous around us and represent the vast storages of low entropy from which irreversible phenomena are fuelled. 

The most common metastable systems in the universe are the large clouds of hydrogen.  Why are they metastable?  Because the protons forming them can fuse into helium, and since this is an irreversible process, helium is a much higher entropy state of its protons than hydrogen.   There are potential barriers for protons to fuse into helium, that make hydrogen a metastable state. But there are processes that can overcome these potential barriers.  A large hydrogen cloud has also a slow gravitational instability that makes it progressively clump \cite{Binney2008}, emitting heat but also increasing pressure and temperature\footnote{Gravitational systems typically have negative heat capacity \cite{Callender2009}.} at its center until the potential barrier preventing hydrogen to fuse into helium becomes insufficient.  Hydrogen starts burning, further increasing temperature, and a channel for rapid increase of entropy is open: a star is born.  

A star like the sun is a strongly irreversible phenomenon.  It produces vast amounts of photons full of free energy, that impact the Earth and fuel a huge amount of irreversible phenomena on the Earth surface, including the entire biosphere.  Hence the entire irreversibility of life can be traced to the low entropy of the initial metastable hydrogen clouds. 

How could the the protons in the hydrogen clouds be in a low-entropy state, if the matter content of the early universe was in thermal equilibrium, as it seems to be, according to standard cosmology \cite{Weinberg2008}?    

The answer is precisely the fact that the expansion of the universe has been too fast for equilibrium to keep up. When the volume was small, hydrogen and helium where in thermal equilibrium, as the standard cosmological model indicates, but then the expansion of the universe became too fast for the long hydrogen-to-helium thermalization time at lower temperatures, and the hydrogen remained trapped in a low-entropy metastable state. This is the low entropy fuelling the majority of the irreversible phenomena we see, including life. 

The thermal history of the universe is therefore very similar to the rapid expansion of the gas in the model discussed above.  As in that model, an initial equilibrium system undergoes a rapid expansion and this generates irreversibility.   

Since the cosmological scale factor $a(t)$ is not manoeuvred by an external force from outside the universe, but is rather a dynamical variable interacting with the rest, the proper model is the second version of the gas with the piston.    As in that model, the initial state was not at equilibrium because while matter was so, there was a single degree of freedom, the cosmological scale factor $a(t)$, badly out of equilibrium.

The energy exchanges between this single degree of freedom and all the others drive the entire irreversibility of the universe we see.   By far the dominant source of the irreversibility we observe is the single fact that the scala factor $a(t)$ was far out of equilibrium in the early universe.   Below I discuss and criticise some alternative interpretations of past low entropy that are common in the literature. Before that, however, let me be a bit more precise.

\subsection{Going out of equilibrium keeping entropy constant}

Consider a co-moving volume of the universe containing two species of matter.  Cal $V$ the volume, $U$ the total internal energy of the matter and $\rho$ the relative density (say the number of particles of the first species over the total number of particles).  The entropy is a function of these macroscopic variables. 
\be
S=S(V,U,\rho).
\ee 
Say at some initial time the value of the macroscopic variables is $(V_{0},U_{0},\rho_{0})$ and the entropy $S_0=S(V_{0},U_{0},\rho_{0})$. Consider the expansion to a state $(V,U,\rho)$.  Because of homogeneity, there is no exchange of energy between co-moving regions, but matter exchanges energy with the gravitational field, therefore in general $U\ne U_0$. Since there is no exchange of heat either, as long as the expansion is reversible, entropy $S$ remains constant: 
\be
S(V,U,\rho)=S_0.
\ee 
In the course of an expansion from the volume $V_0$ to a volume $V$, there are two possibilities: either the two species do not interact, each expanding freely, and therefore $\rho=\rho_0$, or they can be transformed into one another and $\rho$ changes.  

In the first case, the change of $U$ can be computed from the last equation. That is, the final value $U_{free}$ of the internal energy  is determined by 
\be
S(V,U_{free},\rho_0)=S_0. \label{deffree}
\ee 
In the second case, namely if the two species interact and remain in thermal equilibrium, the density changes  adjusting $\rho$ to the value $\rho_{eq}(U,V)$ that maximizes the entropy at given $U$ and $V$. 
\be
\left.\frac{\partial S(V,U,\rho)}{\partial \rho}\right|_{\rho=\rho_{eq}(U,V)}=0.
\ee 
In this case the entropy is only a function of two variables,
\be
S(V,U)=S(V,U,\rho_{eq}(U,V)), \label{S2}
\ee 
and the change of the energy is determined by \emph{this} function remaining constant. That is, if the species interact, the final energy $U_{int}$ is determined by 
\be
S(V,U_{int})=S(V,U_{int},\rho_{eq}(U_{int},V))=S_0. 
\ee 
In general $U_{free}\ne U_{int}$ because the work that the gravitational field does on the matter in the expansion from $V_0$ to $V$ depends on whether there is  interaction between the species or not: in the first case it must also account for energy needed to transform one species in the other. 

Hence the two evolutions take the system to two states with the same volume and the same entropy, but different energy.  

Imagine now that after an evolution where the species were isolated a channel of communication opens that allows one species to transform into the other. Matter internal energy is conserved during this equilibration therefore the state $(V,U_{free},\rho_1)$ evolves to the state $(V,U_{free},\rho_{eq}(V,U_{free}))$. This is different from the state $(V,U_{int},\rho_{eq}(V,U_{int}))$ because $U_{free}$, as we have seen, is different from $U_{int}$, hence its entropy is different from $S_0$. It is higher, because it comes from a transformation towards equilibrium. Therefore there is a net increase of entropy 
\begin{eqnarray}
\Delta S&=& S(V,U_{free})-S(V,U_{int})
\nonumber \\ &=&
 S(V,U_{free})-S_0.
\end{eqnarray}
where $U_{free}$ is defined in \eqref{deffree} and $S(U,V)$ in \eqref{S2}.  

This result may seem strange, because the system goes out of equilibrium even if its entropy remains constant. But of course there is nothing wrong with this: if we slowly shift a wall dividing a box filled with an ideal gas into two parts, the pressure dis-equilibrates, and --in this sense-- we bring the system out of equilibrium without changing its entropy. We simply add energy bringing the system to a state where its maximal entropy can be higher: opening a hole in the wall starts an irreversible process.  Stars are like that opening in the wall.

\subsection{Why other suspects are innocent}

The scale factor $a(t)$ is a component of the gravitational field. It is one of the dynamical variables of gravity.  Therefore the above account shows that gravity has played a key role in the thermal history of the universe, via $a(t)$.   

A second role of gravity in the thermal history of the universe mentioned above is in  the collapse of the hydrogen clouds that ignites stars. 

Roger Penrose has emphasized a different role of gravity for past low entropy, and this idea has had a strong impact on the theoretical physics community, especially on the gravitational community. In the standard cosmological model, the gravitational field is very close to homogeneity and isotropy in the early universe.  Penrose pointed out  that this is an ``extremely special" configuration of the gravitational fields, because a ``generic" configuration of the geometry is strongly crumpled, not homogeneous \cite{Penrose1979}.  This can be seen by studying the evolution of a re-collapsing universe: near the final big crunch, matter has collapsed into a large number of black holes and geometry is highly inhomogeneous. This progressive crumpling of the geometry can be seen as an entropy increase, from a low-entropy initial nearly spatially flat macro-state underpinned by a single microstate, towards a final higher entropy crumbled macro-state. 

All this is of course theoretically correct, but it seems to me that there is clear evidence that this is not the source of the bulk of the irreversibility that we actually observe in our real universe.  The reason is the following. 

Immagine a universe --distinct from ours-- where general relativity was not true, gravitational interaction was instantaneous and Newtonian, and the gravitational field had no local degrees of freedom at all, except for the scale factor.   As fas as we understand, such universe could be very similar to ours, with nearly the same thermal history.  Therefore gravitational waves and metric perturbations of a Friedman cosmology other than the Newtonian potential play a negligible role in the overall thermal history of our universe: if we shut down the degrees of freedom of gravity other than the scale factor, we still have essentially the same large irreversibility.  If this is the case, the irreversibility we see is \emph{not} the consequence of the existence of these other gravitational degrees of freedom, because their absence would not avoid the observed irreversibility. 

I am not disputing Penrose's observation that the initial state of spacetime is peculiar in the phase space of general relativity. But it seems clear that this peculiarity is not the main responsible for the irreversibility we \emph{actually} see around us. The irreversibility we see is pretty much unrelated to the behaviour of gravitational radiative modes, and follows largely from one single degree of freedom, the scale factor, being far from an equilibrium configuration. 

Gravity influences the thermal history of the universe also by giving rise to the clumping of matter that ultimately leads to the rich structure of the universe.  A detailed calculation (see for instance \cite{Wallace2010}), shows that a cloud itself lowers its entropy by shrinking, but emits heat, which rises the external (and the global) entropy.  This suggests that we can relate the initial low entropy fuelling present irreversibility to the initial uniformity of the distribution of matter, whose entropy can increase by gravitational clumping.  

However, the total entropy produced by a star burning in enormously larger than the entropy produced by the heat emitted by the contraction of the hydrogen clouds.  Hence once again, although matter uniformity might marginally contribute to current irreversibility, this effect is negligible with respect to the main one: the irreversibility produced by the fast increase of the scale factor, that have left hydrogen and helium badly out of thermal equilibrium. The universe, in other words, has gone badly out of equilibrium much before any significative beginning of clumping of matter.   

I think we can safely conclude that the past low entropy that gives rise to observed irreversibility is very largely in the single degree of freedom $a(t)$ being far away from an equilibrium value. 

\section{Part II: Why?}

In this second part of the article, I discuss how to interpret the fact that the entropy of the early universe was low.  To star with, let me review our present best understanding of the origin of the second law of thermodynamics. (See for instance \cite{Lebowitz1993} or the general introductory parts in \cite{Price} and  \cite{Albert2000}.)  This allows me to sharpen the question. 

It is easy to make sense of the ubiquitous phenomenon of irreversibility of the observed processes in nature in statistical terms, \emph{if} the initial state of these processes has low entropy.  Indeed, the vast majority of the micro-states that underpin a low-entropy macro-state evolve towards a microstate underpinning a higher entropy configuration.  Therefore, barring extraordinarily atypical micro-states, entropy grows  \emph{if} the initial macro-state of these processes has low entropy.

Given any specific irreversible process, its initial low entropy can be understood because of the way the process was prepared, by us or by Nature.  In either case, preparation of an initial low-entropy state requires that the previous processes giving rise to the preparation themselves started of from an even lower entropy.  

Tracing back in time, we arrive at the low entropy of the early universe.   Therefore current irreversibility depends on early low-entropy in the universe.  The  open question about the arrow of time, therefore, is not why entropy grows: it is why entropy was low to start with.   

This story implies that there is something ironic in the state of our understanding of irreversibility: the statistical understanding of entropy growth is grounded on the idea of \emph{genericity} (generic micro-states underpinning low entropy states evolve towards higher entropy), but to make use of this idea we need to deal with the fact that the initial state had very low entropy, and very low entropy means to be very badly \emph{non-generic}.   Thus, our current understanding of irreversibility is based on an assumption of \emph{genericity} (the micro-states underpinning macro-states are typical) and an assumption of \emph{non-genericity} (initial entropy was very low).  Can we make sense of this? 

A possible attitude towards this question is to discard it.  After all, ``why" questions need to stop at some point.  This is an attitude utilised for instance by David Albert, who recommends to promote the ``past hypothesis", namely the statement that entropy was low in the past, to a sort of law of nature, in the sense of being a general statement from which we can derive predictions that turn out to be true \cite{Albert2000}.   To some extent, the discussion above supports David Albert's position.  Having identified initial low entropy with a small value of the scale factor, allows us to say that all current irreversibility is driven by the initial smallness of the universe, and this is it, as far as the arrow of time is concerned.  

As for any other general fact of nature, not asking further why's is a sensible option. Unless (or until) one finds something better.  Can we find something better?  

\subsection{The role of corse graining}

Let us reconsider the model of the gas coupled to the piston, and assume for simplicity that the total energy is bounded.  Consider the space $\cal M$ of all possible motions of this system. This is a phase space (because at any given time $t$ it is in one-to-one correspondence with the space of the initial data) and carries a natural measure (the Liouville measure $d\mu$, normalized by $\int d\mu=1$, which does not depend on $t$).  For the vast majority of the motions in $\cal M$, and for any give time $t$, the piston is near its equilibrium value.  
\be
                    {\int_{\cal M} X(t)\;d\mu}=X_o.
\ee
If we witness a motion where the piston is far away from this value at some time $t_{0}$, we are witnessing a very atypical motion.  For the same reason, we may say that we are witnessing a very ``atypical" motion of our universe, among the motions allowed by its dynamics. 

But there is something missing in this account of what ``atypical" means.  The reason is that the phase space of any system with many degrees of freedom is very large, therefore \emph{any} single individual motion is very atypical, at the microscopic level.   In fact, pick an arbitrary motion in $\cal M$.  Say a  generic degree of freedom $x^{(n)}$ of the system has mean value
\be
                    {\int_{\cal M} x^{(n)}\!(t)\,d\mu}=x^{(n)}_o.
\ee
But if there are are many degrees of freedom there will most likely be at least one, say $x$, such that $x(t_{0})$ is very different from $x_o$. In other words, in a system with many degrees of freedom there will always be some that are far out of equilibrium.  So, having a \emph{single} degree of freedom that is far out of equilibrium is not ``atypical": to the contrary: it is a very generic occurrence: it is very typical!

What was missing above is the fact that the notion of ``typical" or ``atypical", like the notion of ``low entropy" or ``high entropy" make sense only if we patch together motions in families that we deem indistinguishable, namely if we ``coarse grain" the phase space.  Typicality makes no sense for micro-states alone: it only makes sense if macro-states are defined.  

A macro-state is a defined as a subset $m$ of $\cal M$, or more in general, by a (normalised) distribution function $\rho$ on $\cal M$.  Then the ``typicality" of the macro-state is measured by the size of $m$ 
\be
                   S_q=\log {\int_{m} d\mu}
\ee
or the entropy of $\rho$ 
\be
                   S= -{\int_{{\cal M}} \rho \log \rho}. 
\ee
Macro-states can be defined by picking up some variables of the system and calling them ``macroscopic".  For instance, in the second model considered above, the volume of the cylinder (namely the position of the piston) and the total energy $E$ can be considered ``macroscopic" variables. They defines a macro-state formed, say, by the region ${\cal M}_{V,E}$ in $\cal M$ where at some $t$ the volume is $V$ or smaller and the energy equal or lower than $E$. Its entropy $S_{V,E}$ is a well defined quantity.  If the system is isolated $E$ is conserved and $S_{V,E}$ is maximised by a value $V_o$ of the volume. This is the value that determines the equilibrium position $X_o$ of the piston (in a further expansion of the gas, the increases of the configuration space does not balance the reduction in phase space caused by the energy that has to go in potential energy of the spring.)

The moral is that the reason why we say that a state where $X(t_{0})$ is  far away from $X_o$ is low entropy is not because it is a ``atypical" microstate: any microstate is individually maximally atypical.   Rather: {\em the reason why we say this state is low entropy is because the variable $X$ is special: it is macroscopic}. 

Although rarely emphasized, it is essential to recall that irreversibility is a macroscopic notion.  It is a property of a certain coarse graining.  It is not a property of a microscopic dynamical evolution.  In Appendix \ref{A2} I recall and discuss this essential point. 

Let us translate this to the thermal history of the universe.   We have seen that the initial low entropy is due to the far-from-equilibrium value of $a(t)$ in the early universe.   In a generic high-entropy state, many individual variables are far from their equilibrium value.  The reason why the far-from-equilibrium value of $a(t)$ determines low entropy is because we treat $a(t)$ as a macroscopic variable, not a microscopic one.  

This is the main conceptual point I wish to make in this paper.   Past low entropy is due to the fact that a single variable $a(t)$ that happened to be far from its  equilibrium value in the early universe is a variable that we treat as ``macroscopic".  Why we treat it as macroscopic?

\subsection{Why we treat the scale factor as macroscopic?}

We are tempted to say that among the many degrees of freedom of the universe, the scale factor $a(t)$ is ``obviously" macroscopic, because it is ``big", it can be directly measured, it interacts directly with all other degrees of freedom, or similar reasons that make it ``special".   If this is the right answer, past low entropy is indeed a manifestation of a very atypical state of motion of our universe, among those allowed by the dynamics that we understand:  a very ``special" degree of freedom, the scale factor, was very far from an equilibrium value, a dozen billions years ago.  This atypical fact drives the entire irreversibility of the observed universe.   This may be the end of the story and the present state of our understanding, which perhaps leaves a sense of not having got to the bottom of the story.  

But before buying this conclusion, let us consider the question in general: why do we treat some variables of a system as ``macroscopic" and others as ``microscopic"?  There are different answers in the literature about this question.   These are:
\begin{enumerate}[(i)]
\item {\em External interactions.} The paradigm of a thermodynamical system is a physical system with \emph{many} degrees of freedom $x_n$, acted upon by an agent that can control and measure \emph{a small number} of variables $X_n$ (the thermodynamical variables).  The $X_n$'s are the macroscopic variables that determine the statistical corse graining that yields the definition of entropy. The coarse graining is not arbitrary: it is physically determined by the external interactions of the system.  Thermodynamics describes the macroscopic behaviour of systems \emph{relative to} the given sets of existing physical interactions between the system and the agent measuring it and acting on it.  The entire universe has no ``external" agent acting on it.  What is it then that fixes the relevant macroscopic observables for the universe as a whole? 
\item {\em Heat versus work.} Thermodynamics has developed as the science describing the exchanges of \emph{heat} and \emph{work} between a system and its environment. Both are exchanges of energy: what determines the difference between heat and work?  Intuitively, it is simple: work is a form of mechanical energy.  But so is heat, at the microscopic level.  When we give a macroscopic account of a process, heat is not anymore considered mechanical energy only because the relevant degrees of freedom are not directly accessible.  The distinction between heat and work is therefore subtle: heat refers to the energy in the microscopic variable, while work refers to the energy in the macroscopic variables.  The distinction between heat and work depends \emph{only} on what we call macroscopic. 
\item {\em Averages.} Boltzmann's approach considers a system $S$ formed by a large number $N$ identical copies $s_n$ of a simple system $s$. The prototypical example is a gas formed by many similar molecules. We can then define a distribution $\rho: \sigma\to R^+$ on the phase space $\sigma$ of $s$, which assigns to any region $R\subset \sigma$ the fraction of the molecules that are in states in this region. That is, if the $n$-th molecule is in the state $x_n\in\sigma_n$, 
\be
\int_{R\subset \sigma} \rho = \frac{1}{N} \sum_n \int_{R\subset \sigma_n}\delta_{x_n}.
\ee
Then an observables $o$ of a single molecule defines a macroscopic observable $O$ for the full system, defined by its average under this distribution. 
\be
O = \int_{\sigma}o\rho. 
\ee
This is a powerful tool that exploits the fact that the system is formed by many copies of a single system.  But it can be applied only for systems composed by a large number of identical subsystems.  This is likely not the case with field theory, general relativity, or the entire universe. 
\item{\em Relative entropy.}  Point ({i}) above can be generalized, and stripped of its anthropocentric and subjectivist aspects as follows. If a system $S$ with \emph{many} degrees of freedom $x_n$, interacts with another system $O$ via an interaction hamiltonian that depends on \emph{a small number} of variables $X_n$ of $S$, then this fact defines a corse graining on $S$, determined by considering the variables 
$X_n$ macroscopic.  This defines an entropy for $S$, which is objective but relative to $O$. 
\end{enumerate}
At the light of this general list, why do we consider the scale factor $a(t)$ a macroscopic variable?  The only definition that applies is the last: relative entropy. Let us see how it applies in our case.

It does apply, for the following reason. We are part of the universe. Therefore we belong to a subsystem $O$ of the universe.  We give a macroscopic description of the physical world, which is based on macroscopic quantities $X_n$ we observe, measure, and sometimes act upon.  These macroscopic quantities are determined by the actual physical interactions between the system we belong to and the rest of the universe, namely between $O$ and the rest of $S$. The scale factor $a(t)$ definitely belongs to the set $X_n$ (otherwise we would not do cosmology).  The conclusion is that past low entropy depends on the fact among the relatively few macroscopic variables that determine our own interaction with the rest of the universe there is one that was badly far from equilibrium in the past. 

So far, I think all this is solid.  Let me now take a speculative step. 

The discussion above gives us a second possible way to interpret the atypicality implicit in past low entropy.   What is atypical is not something pertaining to the universe by itself, but to the interacting couple $(S,O)$.  This opens the possibility that what is atypical is $O$, not the state of $S$. 

This is the idea of the perspectival origin of the arrow of time that was put forward in \cite{Rovelli2017c}. It is based on a simple conjecture in statistical mechanics:
\begin{quote} {\em Conjecture:} In a sufficiently complex dynamical system $S$ with sufficiently many interacting degrees of freedom $x_n$, for any generic finite motion there are some subsystems $O$ that interact with the rest of $S$ via interaction variables $X_n$ that define a coarse graining and hence an entropy for the rest of $S$ that is arbitrarily low at one extreme of the motion. 
\end{quote}
If this conjecture is true, as it seems intuitively obvious, then there is nothing a-typical in the fact that we see low past entropy.  It only indicates that we happen to be one example of these subsystems that the conjecture states exist generically.  (On this, see also \cite{arXiv:1804.04147}. The precise relation with the argument in this reference will be studied elsewhere.) 

The reason we happen to be part of one of these peculiar systems is simply that these are the systems constructed in terms of those macroscopic variables for which there is a strong entropy gradient.  And we are the product of entropy gradients. 

When seeing a strongly oriented arrow of time, we are not seeing a property of the microscopic motion of the universe: we are seeing a feature of those special macroscopic variables that made us.  

 If this is the case, the arrow of time is real, but it is perspectival, like are real but perspectival the rotation of the sky or the setting of the sun. 

\section{Conclusions}

In the first part of this note I have argued that the main features of the thermal history of our universe are clear.  The dominant source of the very low-entropy in the past universe is \emph{only} the smallness of the scale factor, which is far from an equilibrium value. 

In the second part of the note I have discussed the extent to which this fact requires an assumption of non-typicality.  I have observed that the low entropy is not due to the fact that one variable is far from equilibrium, but rather to the fact that one of the variable very far from an equilibrium configuration is \emph{also} a variable that we treat as macroscopic.  

I have then observed that the reason for which we consider this variable macroscopic is not completely clear. I see two possibilities. Either the variable is objectively ``special", or it is special because it belongs to the macroscopic variables determined by the peculiar interaction between a physical subsystem to which we belong and the rest of the universe.  

This second possibility opens up the speculative possibility that the arrow of time is perspectival: if a plausible conjecture on statistical mechanics hold, what may be special is not the state of the universe, but rather the set of macroscopic variables we use to describe the macroscopic universe.  The arrow of time might be real, but perspectival, like the rotation of the sky around us, as argued in \cite{Rovelli2017c}. 

\appendix

\section{A toy cosmology}\label{A1}

\begin{figure}[t]
\includegraphics[height=2cm]{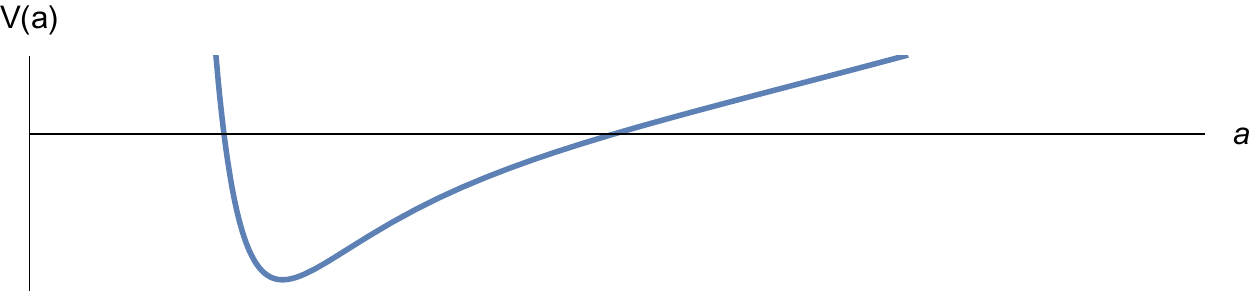}
\caption{Effective potential Eq.\eqref{potent} of the cosmological toy model.}
\label{pot}
\end{figure}

If the discussion above is correct, it should be possible to understand the basics of the thermal history of the universe in terms of a simple toy cosmological model where only uniform matter distribution and the the gravitational scale factor are taken into account.  Let me for extreme simplicity assume that the universe is spatially closed and filled with radiation.  We disregard the cosmological constant, since this has presumably had no effect on the onset of irreversibility.    In this approximation, the cosmological dynamics in proper time $t$ described by the Friedmann equation
\be
\frac{\dot a^2+1}{a^2}=\frac{8\pi G}{3}\rho
\ee
where $\rho$ is the radiation energy density and $\rho^o=\rho a^4$ (the coordinate radiation energy density) is constant.  We write for convenience
\be
\hat\rho\equiv\frac{8\pi G}{3} {\rho}a^4=\frac{8\pi G}{3} \rho_o.
\ee
so that the Friedmann equation reads 
\be
\frac{\dot a^2+1}{a^2}=\frac{\hat\rho}{a^4}
\ee
This universe re-collapses.   Here we are only interested in the initial phases of the expansion, therefore the details of the recollapse are not of interest for us and  we can choose them at wish.  It is convenient to assume a bouncing cosmology, where the recollapse if followed by a bounce and a new expansion, because this situation allows us to control a hypothetical long term evolution.  The dynamics of the bounce can be simply derived  by assuming some negative energy density (of quantum origin) at very small $a$.  We thus modify the  Friedman equation adding a strongly repulsive terms at very short $a$.  I take as  example a term of power -8. Adding this terms, the equation becomes
\be
\frac{\dot a^2+1}{a^2}=\frac{\hat\rho}{a^4}-\frac{c}{a^8}.
\ee
The cosmological dynamics governed by this equation is that of a particle with zero energy moving in the potential 
\be
V(a)=\frac{a^2}{2} \left(\frac{c}{a^8}-\frac{\hat\rho}{a^4}+1\right)\label{potent}
\ee
depicted in Figure \ref{pot}

Since the potential grows for both small and large $a$, the motion is confined in the region where the potential is negative, which is where
\be
\hat\rho-\sqrt{\hat\rho^2-4c}<\frac{2c}{a^4}<\hat\rho+\sqrt{\hat\rho^2-4c}
\ee
which for $c\ll \hat\rho$ gives
\be
\frac{c}{\hat\rho}<a^4<\hat\rho
\ee
The universe bounces back and forth between these two sizes.   Notice that there are solutions only if 
\be
\hat\rho^2>4c.
\ee
The acceleration of the scale factor is governed by the ``force"
\be
F(a)\equiv \ddot a =-\frac{dV}{da}=\frac{a^8 - 3 c + a^4\hat\rho}{a^7}.
\ee
Now let us add something modelling dissipation to this cosmology. Dissipation (apparently) violates the mechanical energy conservation. We can mimic the effect of dissipation by adding small a dissipative friction term to the force, 
\be
F(a)=\frac{a^8 - 3 c + a^4\hat\rho}{a^7}-\beta \dot a
\ee
If $\beta$ is small enough, its effect amount simply to a slow decrease in the energy.  The dynamics is now that of a particle 
slowly lowering the energy in the model, namely by adding a slow time dependent cumulative negative term to the potential (that ``takes away" energy).    The resulting dynamics is clearly that of an universe that oscillates, until slowly setting down to the value of $a$ that minimises the effective potential and maximises the energy.   Here we see a toy example of a universe that has an initial irreversible phase because one of its variables, the scale factor, was out of equilibrium. 

Thos is a  brute manner of modelling dissipation of course, but it serves the illustrative purpose here.  The claim is not that the toy model reproduces the full thermal history of universe (for the moment the data appear to point to a non re-collapsing universe, due to the cosmological constant).  The claim is simply that it is reasonable to interpret the irreversibility of the universe a consequence of the fact that one of its variables, the scale factor, was badly away of any equilibrium value in the early universe.

\section{Irreversibility is a macroscopic notion.}\label{A2}

Irreversibility is a macroscopic notion and \emph{only} a macroscopic notion. Some form of coarse graining is needed even to define irreversibility. I believe that the relevance of this subtle fact for understanding irreversibility is not sufficiently appreciated.  Hence I recall it here in detail. 

Irreversibility is the fact that we witness certain phenomena (a falling glass breaks) and we do not witness their time reversal (glass fragments jump up from the floor and recombine).  But given \emph{any} single microscopic motion of an Avogadro number of atoms, the chances of seeing it are ridiculously small.  Therefore the chances of seeing the time reversal of \emph{any} microscopic evolution that we have witnessed are \emph{always} negligible.  Hence there is nothing surprising in the fact that we  never see the time reversed evolution of a \emph{microscopic} motion that has happened. Even the chance of seeing the \emph{same} microscopic motion twice are ridiculously negligible!
  There is no irreversibility in microphysics. 

How come then that do not see glasses recombine, but we do repeatedly witness glasses breaking?  

The answer is that these are statements about groups of states or groups of motions lumped together, not about individual stated.  A ``glass breaking" is not a single microscopic motion, it is an ensemble of a huge number $N_{forth}$ of possible microscopic motions.   ``Glass fragments recombining" is also an ensemble of a huge number $N_{back}$ of possible macroscopic motions. Since microphysics is time reversal invariant, $N_{forth}=N_{back}$.   

But the number $N_{broken}$ of micro-states that we call a ``broken glass" is far larger than the number $N_{intact}$ of micro-states that we call a ``falling intact glass".  Now, \emph{all} the $N_{intact}$ microstate evolve (back or forth in time does not matter) into one of the $N_{broken}$ microstates, but only a tiny fraction of the $N_{broken}$ microstate can evolve (back or forth in time) into one of the $N_{intact}$ ones. That is: $N_{forth}=N_{back}=N_{intact}\ll N_{broken}$.   The arrow of time is then the following phenomenon: it is an observed fact of nature that the \emph{past} (but not the future) micro-state of the world belongs to the very small, non-generic, set of micro-states in $N_{intact}$.   The notion of ``intact falling glass" captures a proportion of initial states far smaller than what genericity would suggest.  Irreversibility is not a property of a microscopic evolution: it is a property of the lumping notions of ``falling intact glass": \emph{this} is a property of the world which is so un-generic that, it it holds at sometime it would generically be lost at other times. 

This is just another way to say that the ground of irreversibility is past low entropy; but I have spelled it out in details to emphasise the fact that the notion of irreversibility pertains only to macro-states and is meaningless for micro-states.  Without lumping, there is nothing irreversible going on.  Irreversibility is a property of a certain lumping. 

Since this point is essential, let me stress it with one additional example.  Consider a set of balls bouncing on a billiard table without friction between a time $t_a$ and a time $t_b$.  Given an arbitrary motion of the balls during this time span, call ``$a$-balls" those that happen to be in the left side of the table at $t_a$ and ``$b$-balls" those that happen to be in the left side of the table at $t_b$.  Define two macroscopic observables, $O_a$ and $O_b$ as the number of ``$a$-balls", respectively ``$b$-balls", in the left side of the table. $O_a$ and $O_b$ are macroscopical observables and define a coarse graining, hence an entropy.   A moment of reflection shows that entropy, so defined, generically \emph{increases} from $t_a$ to $t_b$ if it is defined by $O_a$, but \emph{decreases} if it is defined by $O_b$. 

Therefore the proper statement about the irreversibility of our universe is not a statement about the evolution of the universe microstate.  It is a statement about a certain coarse graining: we do describe the world by means of a coarse graining and a set of macroscopical observables that define an entropy that was low in the past.   The relevant question is then: what determines which observables of a system are ``macroscopic"?
%\centerline{---}

%I thank. 

%\vfill

\bibliographystyle{utcaps}
\bibliography{/Users/carlorovelli/Documents/library}
\end{document}